\documentclass[%
 reprint,
superscriptaddress,
 amsmath,amssymb,
 aps,
pra,
]{revtex4-2}
\usepackage[T1]{fontenc}
\usepackage{graphicx}
\usepackage{amsfonts}
\usepackage{dcolumn}
\usepackage{bm}
\usepackage{braket}
\usepackage{booktabs}
\usepackage{xcolor}
\usepackage{siunitx,xfrac}
\usepackage[normalem]{ulem}  
\DeclareSIUnit{\sqrthz}{\sqrt{\hertz}}

\newcommand{\weg}{\omega_{eg}}

\begin{document}

\preprint{APS/123-QED}

\title{Linear and continuous variable spin-wave processing using a cavity-coupled atomic ensemble}

\author{Kevin C. Cox}
\email[Corresponding author: ]{kevin.c.cox29.civ@army.mil}
\affiliation{
 DEVCOM Army Research Laboratory, Adelphi, MD 20783 USA
}%

\author{Przemyslaw Bienias}
    \affiliation{Joint Quantum Institute and Joint Center for Quantum Information and Compute Science, NIST/University of Maryland, College Park, Maryland 20742, USA}
    
\author{David H. Meyer}
    \affiliation{
 DEVCOM Army Research Laboratory, Adelphi, MD 20783 USA
}%

\author{Donald P. Fahey}
    \affiliation{
 DEVCOM Army Research Laboratory, Adelphi, MD 20783 USA
}%

\author{Paul D. Kunz}
\affiliation{
 DEVCOM Army Research Laboratory, Adelphi, MD 20783 USA
}%
\author{Alexey V. Gorshkov}
\affiliation{Joint Quantum Institute and Joint Center for Quantum Information and Compute Science, NIST/University of Maryland, College Park, Maryland 20742, USA}

\date{\today}

\begin{abstract}
Spin-wave excitations in ensembles of atoms are gaining attention as a quantum information resource.  However, current techniques with atomic spin waves do not achieve universal quantum information processing.   We conduct a theoretical analysis of methods to create a high-capacity universal quantum processor and network node using an ensemble of laser-cooled atoms, trapped in a one-dimensional periodic potential and coupled to a ring cavity.  We describe how to establish linear quantum processing using a lambda-scheme in a rubidium-atom system, calculate the expected experimental operational fidelities.  Second, we derive an efficient method to achieve linear controllability with a single ensemble of atoms, rather than two-ensembles as proposed in [K. C. Cox et al. ``Spin-Wave Quantum Computing with Atoms in a Single-Mode Cavity'', preprint 2021].  Finally, we propose to use the spin-wave processor for continuous-variable quantum information processing and present a scheme to generate large dual-rail cluster states useful for deterministic computing.
\end{abstract}

\maketitle

\section{Introduction}

Laser-cooled atoms in optical resonators are a building block for many of the most exquisite demonstrations of quantum electrodynamics.  Atom-cavity systems are the basis for state-of-the-art quantum simulators \cite{baumann_dicke_2010,clark_observation_2020}, quantum memories \cite{yang_efficient_2016}, and entanglement-enhanced atomic clocks \cite{malia_free_2020, pedrozo-penafiel_entanglement-enhanced_2020}.  With atom number $N$ commonly between $10^3$ and $10^6$, an ensemble's intrinsic capacity to store quantum information is enormous, with a state space of dimension $2^N$.  Designing quantum platforms that are able to access and process this large amount of quantum information is a grand challenge in atomic science.  Here we analyze a method to store quantum information as collective spin-wave excitations and realize universal quantum computation in a system where the spin waves may be efficiently retrieved into a single optical cavity mode.

Recent experiments have introduced a path to use collective spin-wave excitations, holographically multiplexed, to achieve high-capacity quantum memories \cite{simon_single-photon_2007,parniak_wavevector_2017,cox_spin-wave_2019,vernaz-gris_highly-efficient_2018, heller_cold-atom_2020}, but these experiments have not introduced a method to achieve full linear controllability of spin-wave excitations, a prerequisite for a universal quantum processor.  Spin-wave quantum systems are being realized in multiple physical platforms including atomic vapors \cite{parniak_wavevector_2017,cox_spin-wave_2019,vernaz-gris_highly-efficient_2018, heller_cold-atom_2020}, solid-state crystals \cite{afzelius_demonstration_2010, kutluer_time_2019,lago-rivera_telecom-heralded_2021}, and superconducting circuits \cite{carusotto_photonic_2020}.  Proposals for spin-wave readout of atomic arrays have also been developed \cite{saffman_entangling_2005,muschik_quantum_2008,grankin_free-space_2018}.  But demonstrating platforms that simultaneously achieve universal quantum processing and efficient memory readout is still an outstanding challenge.  In a joint Letter publication \cite{cox_spin-wave_2021}, we present a general scheme for universal linear-optical quantum processing using spin-wave excitations coupled to a single optical mode.  In this Article, we build upon that work by describing, in detail, the experimental methods and performance required to physically realize universal quantum information processing with laser-cooled atoms inside of an optical ring cavity.  

First, we describe a two-ensemble experimental method to realize universal spin-wave quantum processing in an ensemble of alkali atoms coupled to a bow-tie ring cavity and discuss the physical operations required.  Second, we present a general proof of linear controllability using only position-space and momentum-space phase shifts with a single atomic ensemble. We derive the exact two-mode beamsplitters in this one-ensemble scenario, that can be accomplished in constant time even when using a large number of spin-wave modes.  Further, we calculate the expected leading operational errors due to experimental imperfections, and the time required to probabilistically initialize single photons in the system using heralded quantum memory initialization.  Last, we calculate the amount of multi-mode spin-squeezing that may be achieved in the system and how this multi-mode squeezing would perform in a dual-rail cluster state.  We calculate that it may be possible to achieve a continuous-variable cluster state with thousands of modes and greater than $20$~dB of squeezing per mode, a significant computational resource for deterministic continuous variable processing \cite{menicucci_universal_2006, gu_quantum_2009, bourassa_blueprint_2021}.  

\section{Apparatus}
\begin{figure}[t]
\centering
\includegraphics[width=\columnwidth]{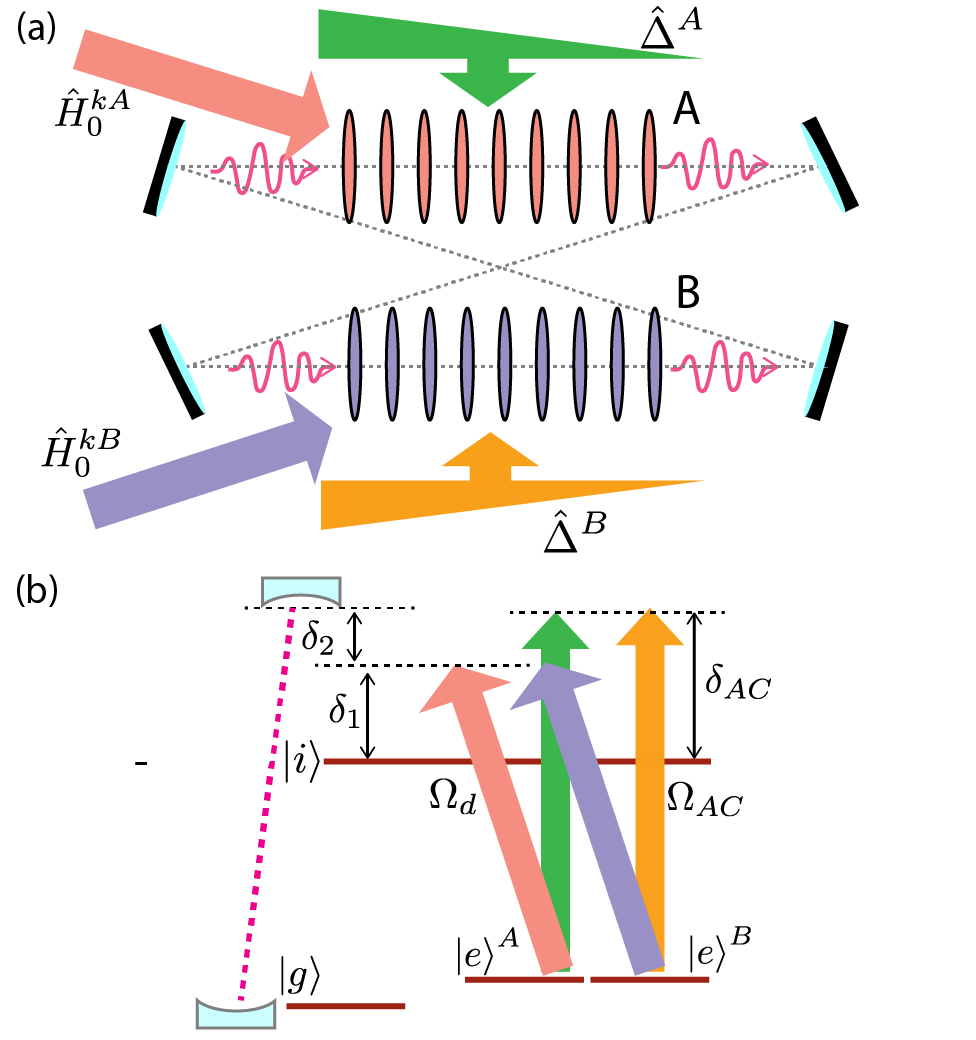}
\caption{Apparatus and level diagram using alkali atoms. (a) Two ensembles are coupled to a single running-wave cavity, and operations are applied using potential gradients (green and orange) with operators $\hat{\Delta}^A$ and $\hat{\Delta}^B$ and Raman dressing beams (red and violet) with associated Hamiltonians $\hat{H}^{kA}_0$ and $\hat{H}^{kB}_0$.  (b) Level diagram.  Raman cavity coupling is controlled with the dressing beams (red and violet).    Optical gradients are applied with large detuning $\delta_{AC}$. The $\ket{e}$ states in ensembles $A$ and $B$ are drawn separately to delineate the ensemble-specific beams.} \label{fig:expt}
\end{figure}

The simplified apparatus and atomic level diagram are displayed in Fig.~\ref{fig:expt}(a) and (b).  We first consider two ensembles, each with $N$ laser-cooled alkali atoms, approximated as three-level atoms with long-lifetime states $\ket{g}$, $\ket{e}$ and optically excited state $\ket{i}$ with linewidth $\Gamma$.  The atoms are confined in two one-dimensional periodic potentials, each with $M$ sites, inside of a running-wave optical cavity.  The two arrays, labelled $A$ and $B$, have the same cavity couplings and experimental parameters.  When discussing parameters and operators that specifically refer to one array or the other, we will denote them with a corresponding superscript A or B.  

A running-wave cavity is required for this experiment in order to distinguish between excitations with left or right-travelling photons.  The arrays are optically interrogated using Raman dressing beams (red and violet in Fig.~\ref{fig:expt}) that stimulate two-photon Raman transitions.  Potential gradients (green and orange in Fig.~\ref{fig:expt}), using laser beams with an intensity variation, are applied perpendicular to the cavity axis.  The off-resonant optical fields are applied with detuning $\delta_{AC}$ and a spatially varying Rabi frequency $\Omega_{AC}(x)$. 

The optical cavity is described by its finesse $f$, full-width half-maximum (FWHM) linewidth $\kappa$, Jaynes-Cummings coupling parameter $g$ associated with the $\ket{g}$-$\ket{i}$ transition and single-atom cooperativity parameter $C = 4 g^2/\kappa \Gamma$.  We consider an apparatus with $C$ near or less than 1, but large collective cooperativity $N C\gg1$.  Atoms in $\ket{e}$ can be made to interact with the cavity mode by applying Raman dressing lasers (red and violet in Fig.~\ref{fig:expt}) with Rabi frequency $\Omega_d$.

With large detuning $\delta_1$ ($|\delta_1| \gg |\delta_2|$)  between the cavity and the $\ket{g}$ to $\ket{i}$ transition ($|\delta_1| \gg \sqrt{N}g , \Omega_d$), the theoretical treatment of Fig. \ref{fig:expt}(a) may be simplified by adiabatically eliminating $\ket{i}$, creating an effective two-level system with dressed excited state $\ket{e'}$ and two-photon scattering rate
\begin{equation}
    \Gamma_2(t) = \Gamma\frac{\Omega_d^2(t)}{4 \delta_1^2},
\end{equation} 
where $\Omega_d(t)$ is the Rabi frequency of the Raman dressing laser.  This results in a new two-photon Jaynes-Cummings coupling parameter 
\begin{equation}\label{eq:2Photon}
    g_2(t) = \frac{g \Omega_d(t)}{2 \delta_1}.
\end{equation}
In addition to the two-photon transition rates, the Raman dressing laser gives rise to shifts in the cavity resonance frequency and the two-photon transition frequency, that must be taken into account (see Sec.~\ref{sec:challenge}).

\begin{figure}[t]
\centering
\includegraphics[width=\columnwidth]{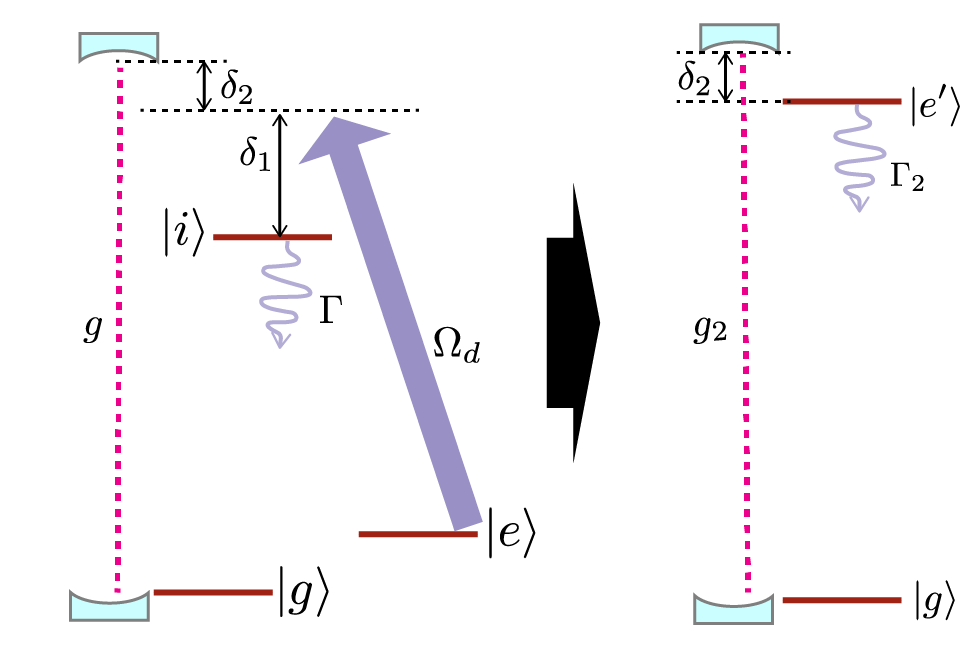}
\caption{Three-level diagram and adiabatic elimination.  The experimental three-level system is treated as a dynamic two-level system with cavity coupling rate $g_2$, excited state lifetime $\Gamma_2$, and atom-cavity detuning $\delta_2$.  } \label{fig:2lev}
\end{figure}

 A diagram that defines the experimental parameters and the relationship between the three-level system and the two-level model is shown in Fig.~\ref{fig:2lev}.  There are several important advantages of this three-level scheme.  Critically, the effective cavity coupling $g_2(t)$ and free space scattering rate $\Gamma_2(t)$ are dynamic, and may be turned on and off at high speed via the dressing laser intensity, proportional to $|\Omega_d(t)|^2$.

\section{Momentum Basis with Alkali Atoms in Cavity}  
As described in the joint Letter \cite{cox_spin-wave_2021}, we can demonstrate linear controllability by defining a set of orthogonal spin-wave modes in the momentum basis.  In Ref.~\cite{cox_spin-wave_2021}, we focused solely on a two-ensemble apparatus where  qubits are arranged in two banks of one-dimensional arrays.  Here, we first discuss how this proposal may be realized in a three-level Raman system with laser-cooled and trapped alkali atoms.  Then, in Section \ref{sec:oneEnsemble}, we show how to extend the proposal to using only a single one-dimensional array.  Using cold atoms, two arrays may be created via counter-propagating trap beams within the cavity,  or the single sites may be created by projecting an additional trapping potential transverse to the cavity mode.  Such a scheme may be useful to precisely control the trap dimension and spacing of the array sites.  

Atomic excitations in the array are described by site-specific lowering 
operators
\begin{equation}
    \hat{a}_x = \frac{1}{\sqrt{n}} \sum_{l=0}^{n-1} \ket{g_l}\!\bra{e_l}.
\end{equation}

The corresponding collective lowering operators in momentum space are
\begin{equation}
    \hat{b}_k = \frac{1}{\sqrt{M}} \sum_{x=0}^{M-1} e^{i 2 \pi k x / M} \hat{a}_x,
\end{equation}
where $k$ can take integer values from 0 and $M-1$.  By initializing and reading out excitations in the modes defined by the $\hat{b}_k$ operators, we ensure that single-site resolution is unnecessary, and all excitations can be tuned into full coupling with the single cavity readout mode \cite{cox_spin-wave_2021}.  

The optical cavity and the dressing laser define a unique momentum $\vec{k}_0$ that can interact with the cavity mode at a given time.  $\vec{k}_0$ is dictated by the microwave qubit frequency $\weg$ and the angle of incidence of the dressing beam relative to the cavity mode \cite{simon_interfacing_2007}.

The set of collective excitations with spin-wave momentum $\vec{k} \in \vec{k}_0 + \{0,1,2...M-1\} \hat{x}/x_0 $ is the orthogonal set of momenta that we consider, where $x_0$ is the array spacing.  Subsequently, we leave the momentum offsets as implicit, and simply label the momentum by an integer $k \in \{0,1,2...M-1\}$. The orthogonality condition between spin-wave modes $\hat{b}_k$ is critical to universal computing, since it implies that collective excitations in mode $\hat{b}_k$, for example, are forbidden to emit into or interact with the cavity mode if $k \neq 0$. 

In the limit of large atom number per site $n$ and low excitation number, the collective spin raising and lowering operators are directly analogous to harmonic oscillator operators, and the cavity-ensemble system behaves as a system of coupled harmonic oscillators. This well-established limit is known as the Holstein-Primakoff approximation. The coupling strength between the ensemble and cavity is then given by the collective vacuum Rabi splitting, $\Omega_2 =2 g_2 \sqrt{N} $.

\subsection{Operations}

The Raman dressing interrogation beams (red and violet) and AC Stark shift gradient beams (green and orange) shown in Fig.~\ref{fig:expt} define two Hamiltonians that may be applied to either ensemble A or B (denoted by superscripts when necessary).  These operations lead to the dynamics described in detail in Ref.~\cite{cox_spin-wave_2021}.  Here we briefly review the Hamiltonians and the slight modifications that come from operating in a Raman system.  The AC Stark shift gradient Hamiltonian in the Raman system is
\begin{equation}\label{eq:prastark}
    \hat{H}_\Delta = \sum_{x=0}^{M-1} \frac {\hbar \Omega_{AC}^2(x)}{4 \delta_{AC}} \hat{a}^\dagger_x\!\hat{a}_x.
\end{equation}
With a choice of $\Omega^2_{AC}(x) \propto x$, this Hamiltonian allows arbitrary shifts in momentum $k^A$ and $k^B$, and has a corresponding unitary that we denote $\hat{\Delta}$:

\begin{equation}
    \hat{\Delta} = \sum_{x=0}^{M-1} \left[\sum_{l=1}^n\left(\ket{g_l}\bra{g_l}+ e^{2 \pi i x/M} \ket{e_l}  \bra{e_l}\right)\right].
\end{equation}
The behavior of $\hat{\Delta}$ is similar to other gradient quantum memories using collective ensembles \cite{hosseini_unconditional_2011, ledingham_solid_2015}.  Equation \ref{eq:prastark} differs from Eq.~2 in the joint Letter \cite{cox_spin-wave_2021} due to the Raman configuration. 

In this proposal, atom-cavity interactions are governed by the Raman dressing lasers. We assume that $\delta_2 \gg \Omega_2$, so that  the collective atom-cavity interaction is dispersive, with Hamiltonian
\begin{equation} \label{eq:k0Ham}
    \hat{H}^k_0 =  -\frac{\hbar\Omega_2^2}{4 \delta_2}\hat{b}^\dagger_0 \hat{b}_0,
\end{equation}
where $\Omega_2$ and $\delta_2$ are the effective coupling strength and detuning for the two-photon transition [Eq.~(\ref{eq:2Photon})].

When $\hat{H}^k_0$ is applied simultaneously to ensemble $A$ and ensemble $B$, a spin-wave beamsplitter Hamiltonian results,
\begin{equation}
\begin{split}
\hat{H}^{BS} &=  -\frac{\hbar \Omega_2^2}{4 \delta_2} (\hat{b}^{\dagger A}_0 + \hat{b}^{\dagger B}_0)(\hat{b}^A_0 + \hat{b}^B_0)\\ 
& = a \begin{pmatrix} 1 & 1\\1 & 1 \end{pmatrix},
\end{split}
\end{equation}
for $a \equiv - \hbar \Omega_2^2/(4 \delta_2)$, where the second line has been written in the $(\ket{b^A_0},\ket{b^B_0})$ basis.  $\hat{H}^{BS}$ can be verified as a beamsplitter Hamiltonian by again calculating the unitary evolution $\hat{\Pi} = e^{-i \hat{H}^{BS} t/\hbar}$, which can be written as  
\begin{equation}
    \hat{\Pi} =  \frac{1}{2}\begin{pmatrix} 1 + 1 e^{-2 i t a}  &-1 + 1 e^{-2 i t a}\\-1 + 1 e^{-2 i t a}&1 + 1 e^{-2 i t a} \end{pmatrix}.
\end{equation}

This A-B beamsplitter, the translation operators $\hat{\Delta}$, and mode phase shifts from Hamiltonian $\hat{H}^k_0$, together allow one to create an arbitrary linear unitary in the $2M$ mode system, a capability we refer to as linear controllability.  Linear controllability is the precise requirement for linear-optical quantum computing, as discussed in the joint Letter \cite{cox_spin-wave_2021}.  Next, we discuss a general proof of linear controllability that does not require two separate ensembles, and discuss alternative beamsplitter constructions.

\section{Quantum Processing with One Ensemble}\label{sec:oneEnsemble}

\subsection{Controllability proof}
So far in this work, we have focused on using two atomic ensembles and performing beamsplitters between spin waves in each.  This is a simple experimental realization, but it is not fundamentally necessary for linear controllability. We now present a general proof that phase shifts alone are sufficient for controllability, even with one ensemble.   For the general proof, we allow ourselves to utilize Hamiltonian generators $\ket{a_x}\bra{a_x}$, with $x = 0, \dots, M-1$, and $\ket{b_0}\bra{b_0}$. These generators correspond to the applications of phase shifts in position space and momentum space, that don't involve single site addressing.

A necessary and sufficient condition for controllability on the underlying $M$-dimensional Hilbert space is that our $M+1$ Hamiltonian terms generate the $(M^2-1)$-dimensional Lie algebra $\mathfrak{su}(M)$ \cite{jurdjevic_control_1972,brockett_lie_1973}. Working in the $\ket{a_x}$ basis, we first construct all the $M-1$  diagonal generators by taking linear combinations of $\ket{a_x}\bra{a_x}$. We construct half (i.e.\ $M (M-1)/2$) of all the off-diagonal generators by considering, for $j \neq l$, 
\begin{align} \label{eq:control0}
&[[\ket{b_0}\bra{b_0},\ket{a_j}\bra{a_j}],\ket{a_l}\bra{a_l}] \nonumber \\
&\propto \frac{1}{\sqrt{M}} [\ket{b_0}\bra{a_l}-\ket{a_l}\bra{b_0},\ket{a_j}\bra{a_j}] \nonumber \\ &\propto -\frac{1}{M} 
(\ket{a_l}\bra{a_j}+\ket{a_j}\bra{a_l}).
\end{align}
We construct the remaining $M (M-1)/2$  off-diagonal generators by considering, 
\begin{align}
& \frac{1}{M}[\ket{a_l}\bra{a_j}+\ket{a_j}\bra{a_l},\ket{a_l}\bra{a_l}] \nonumber \\
&= \frac{1}{M}(i \ket{a_j}\bra{a_l}- i \ket{a_l}\bra{a_j}).
\label{eq:control}
\end{align}
The generators synthesized in Eqs.\ (\ref{eq:control0},\ref{eq:control}) are precisely the off-diagonal beamsplitter generators that are not typically accessible in spin-wave quantum memories.  Note that we have only used the $\ket{b_0}$ momentum-space phase shift.  However, the off-diagonal elements are reduced by a factor of $1/M$.  For this reason, although the $\hat{H}^k_0 \sim \ket{b_0}\!\bra{b_0}$ phase shift generator is sufficient for controllability, the beamsplitter interactions become weaker as the system size $M$ grows.  Accomplishing arbitrary unitary dynamics would require pulse sequences that grow unfavorably with $M$.  Next, we present a different Hamiltonian generator that allows us to implement spin-wave beamsplitters in a single ensemble that alleviates this deleterious scaling, showing that arbitrary two-mode beamsplitters can be implemented in constant time, even in the large-$M$ limit.  

\subsection{Numerically optimized beamsplitters}
Since controllability is possible without two ensembles in principle, it is worthwhile to describe a construction that achieves efficient linear controllability in a single ensemble. Unlike in the previous section, we will work here in the momentum basis $\ket{b_k}$.  A beamsplitter between spin-wave modes $\hat{b}_j$ and $\hat{b}_l$ can be generated by the Hamiltonian
\begin{align}
    \hat{H}^{BS}_{jl} &\propto (\ket{b_j} + \ket{b_l})(\bra{b_j} + \bra{b_l}),
\end{align}
 with $j \neq l$.  Expanding out the state $\ket{b_j} + \ket{b_l}$ shows that this Hamiltonian corresponds to both phase modulation and amplitude modulation across the spin wave:
\begin{align}
    \ket{b_j} + \ket{b_l}&= \frac{1}{\sqrt{M}}\sum_{x=0}^{M-1} (e^{2 \pi  i j x} + e^{2 \pi i l x}  ) \ket{a_x} 
\end{align}
due to the summation of the complex amplitudes at each site.  For this reason, we are not able to apply this Hamiltonian directly using only phase shifts.  This is one reason that previous experiments have not achieved complete controllability in a spin-wave register.

In order to overcome this challenge, we propose to implement a similar Hamiltonian, that is generated using only phase shifts but nonetheless yields efficient unitary controllability.  The modified Hamiltonian is
\begin{align}
    \hat{H}'_{jl} &\propto  \ket{b'}\bra{b'}, \\
    \ket{b'} &= \frac{1}{\sqrt{M}} \sum_{x=1}^{M} \operatorname{Exp}\left[i \operatorname{Arg}(e^{2 \pi  i j x/M} + e^{2 \pi i l x/M}  )\right] \ket{a_x}. 
\end{align}
$\hat{H}'_{jl}$ only applies the phase component of the beamsplitter Hamiltonian $\hat{H}^{BS}_{jl}$.  This Hamiltonian may be constructed using only phase shifts, by turning on the cavity coupling Hamiltonian $\hat{H}^k_0$ [Eq.~(\ref{eq:k0Ham})] to a spin-wave state with the nontrivial phase $\operatorname{Arg}(e^{2 \pi  i j x/M} + e^{2 \pi i l x/M})$, instead of the $k=0$ mode.  Experimentally this would be done in a two-step process, first applying the phase modulation, and then turning on the cavity coupling.  The modified beamsplitter Hamiltonian $\hat{H}'_{jl}$ does not generate an exact two-mode beamsplitter on its own.  However, using $\hat{H}'_{jl}$ in conjunction with the two other available Hamiltonians $\hat{H}_j \propto \ket{b_j}\bra{b_j}$ and $\hat{H}_l \propto \ket{b_l}\bra{b_l}$ in a multi-pulse sequence allows us to do so.  Next, we present the procedure to numerically and analytically generate a precise two-mode beamsplitter using $\hat{H}'_{jl}$.

The three operators $\hat{H}'_{jl}$, $\hat{H}_j$ and $\hat{H}_l$ define a three-level system with basis states $\ket{b_j}$, $\ket{b_l}$, and $\ket{b^*}$.   $\ket{b^*}$ is defined so that $\ket{b^*}$, $\ket{b_l}$, and $\ket{b_j}$ form an orthonormal basis of the three-dimensional space spanned by $\ket{b'}$, $\ket{b_j}$, and $\ket{b_l}$. To construct $\ket{b^*}$, we substract from $\ket{b'}$ its projections on $\ket{b_j}$ and $\ket{b_l}$ and normalize the result.

\begin{figure}[t]
\centering
\includegraphics[width=\columnwidth]{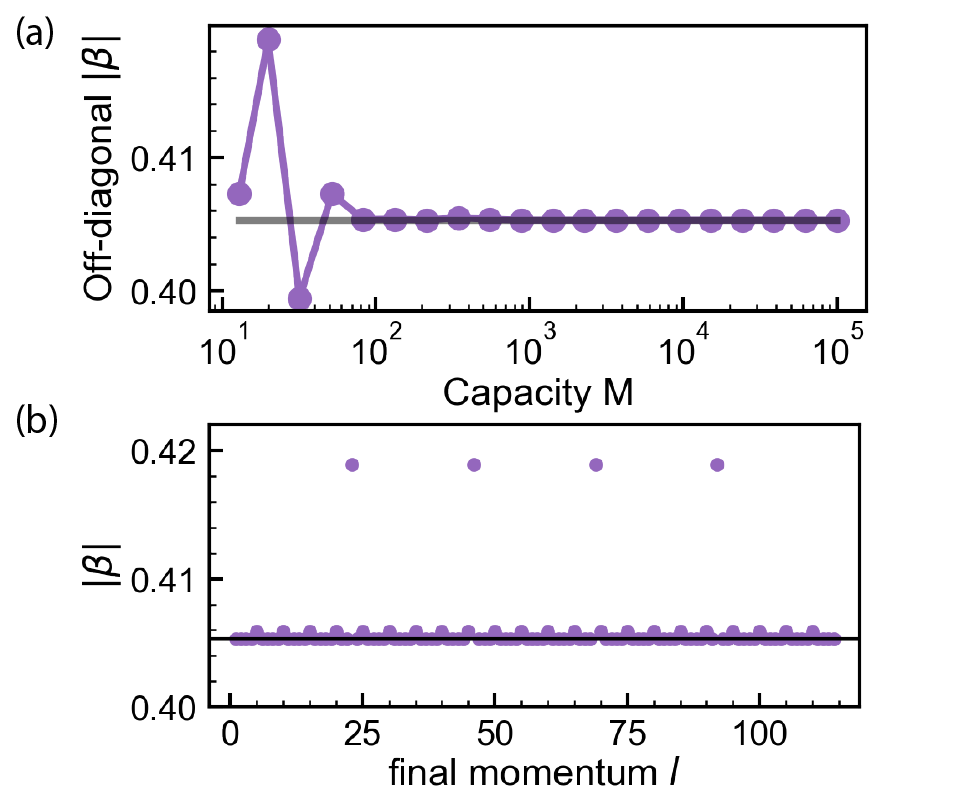}
\caption{Beamsplitters with a single ensemble.  (a) The off-diagonal matrix element $\beta$ is plotted for the optimized momentum-space beamsplitter $\hat{H}'_{jl}$.  $\beta$ remains large for all values of $M$ indicating an effective beamsplitter, and approaches a constant value $4/\pi^2$ for large $M$ (solid line).  (b) $\beta$ is plotted for $M = 115$ and $j=0$ versus $l$, indicating that beamsplitters are possible for all values of $l-j$.  Slight variation in $\beta$ is observed for values of $l-j$ that are a large divisor of $M$, evident in the plot for values of $l$ that are multiples of 5 and 23. } \label{fig:Numeric}
\end{figure}

Efficiently generating a beamsplitter requires the generating Hamiltonian $\hat{H}'_{jl}$ to have large off-diagonal element $\beta = \bra{b_l} \hat{H}'_{jl} \ket{b_j}$.  We write $\hat{H}'_{jl}$ in the basis $(\ket{b_j}, \ket{b_l}, \ket{b^*})$:
\begin{align}
    \hat{H}'_{lj} &= \begin{pmatrix} \alpha & \beta^* & \gamma^* \\ \beta & \epsilon & \zeta^* \\ \gamma & \zeta  & \theta \end{pmatrix}.
\end{align}
The element $\beta$ describes the beamsplitter strength.  In Fig.~\ref{fig:Numeric}~(a), we plot the magnitude of $\beta$ for the $\hat{H}'_{jl}$ Hamiltonian (purple) as a function of $M$ for$j = 1$ and $l = 8$.  This plot shows that  $\hat{H}'_{jl}$ can be used to generate an effective beamsplitter at large $M$, since $\beta$ remains at a value of nearly 0.4.  Figure \ref{fig:Numeric}~(b) displays the magnitude of $\beta$ for $M=115$ as a function of $l$ for $j = 1$.  The exact values of the elements of $\hat{H}'_{jl}$ depend on $j$, $l$, and $M$, but critically, they remain large for all values, and approach the value of $4/\pi^2$ for large $M$, shown as a solid dark line in Fig.\ref{fig:Numeric}(a) and (b). The resulting value of $4/\pi^2$ is derived by calculating $\braket{b_l|b'}$ (or equivalently $\braket{b_j|b'}$) for large $M$.  For values where $l-j$ are a multiple of a large divisor of $M$, departures from the nominal value $\beta = 4/\pi^2$ are observed.  For example, small deviations at multiples of 5 and 23 can be observed in Fig. \ref{fig:Numeric} for $M = 115$ where 23 and 5 are the only nontrivial divisors of 115. 

For large $M$, when the value of $l-j$ is not a large integer divisor of $M$, the Hamiltonian $\hat{H}'_{lj}$ becomes,
\begin{align}
    \hat{H}'_{lj} &= \begin{pmatrix} 4/\pi^2 & 4/\pi^2 & \sqrt{\frac{4}{\pi^2}(1-\frac{8}{\pi^2})} \\ 4/\pi^2 & 4/\pi^2 & \sqrt{\frac{4}{\pi^2}(1-\frac{8}{\pi^2})}\\ \sqrt{\frac{4}{\pi^2}(1-\frac{8}{\pi^2})}& \sqrt{\frac{4}{\pi^2}(1-\frac{8}{\pi^2})}  & 1-\frac{8}{\pi^2} \end{pmatrix}.
\end{align}
Importantly, in this limit,  $\hat{H}'_{lj}$ becomes independent of $l$, $j$, and $M$.

To generate exact beamsplitters, we numerically optimize amplitudes in an interleaved pulse sequence.  The desired 50-50 beamsplitter unitary is 
\begin{align}
    \hat{U}^{BS} = \frac{1}{\sqrt{2}} \begin{pmatrix} 1  &  i & 0\\  i & 1 & 0 \\ 0 & 0 & \sqrt{2} \end{pmatrix}.
\end{align}
For large $M$, we find that $\hat{U}^{BS}$ can be achieved in a seven-pulse sequence of the form
\begin{equation}
\begin{split}\label{eq:BS}
    \hat{U}^{BS} &= \hat{U}'_{lj}(\theta_7) \cdot \hat{U}_{l}(\theta_6) \cdot\hat{U}_{j}(\theta_5)\\ &\quad\cdot \hat{U}'_{lj}(\theta_4) \cdot \hat{U}_{l}(\theta_3) \cdot \hat{U}_{j}(\theta_2) \cdot \hat{U}'_{lj}(\theta_1),
\end{split}
\end{equation}
where each unitary is derived from it's respective Hamiltonian: $\hat{U}_{l}(\theta) = e^{-i\hat{H}_l \theta}$, $\hat{U}_{j}(\theta) = e^{-i\hat{H}_j \theta}$ and $\hat{U}'_{lj}(\theta) = e^{-i\hat{H}'_{lj}\theta}$.  The solutions for the rotation angles $\{\theta_1,...\theta_7\}$ are given analytically in a supplemental file \footnote{A pdf printout of a Mathematica notebook that calculates these values is included as Supplementary Material, available online.}.  The approximate numerical values are shown below. This solution is valid for arbitrary values of $j$ and $l$.
\begin{table}[h]
\begin{tabular}{cc} 
 \toprule
 $i$ & $\theta_i/2\pi$\\
 \midrule
 $1$ & 0.347136 \\
 $2$ & 0.222136 \\ 
 $3$ & 0.222136\\ 
 $4$ &  0.125\\ 
 $5$ & 0.652864\\ 
 $6$ & 0.652864\\ 
 $7$ & 0.777864\\ \bottomrule
\end{tabular}
\label{tab:angles}
\caption{Approximate numeric rotation angles for each unitary of the optimized 50-50 beamsplitter in Eq.~\ref{eq:BS}.}
\end{table}

The key achievement of this construction is that exact two-mode beamsplitters may be achieved between any two spin-wave modes in constant (independent of $M$) time, even for large $M$.  This type of connectivity is unique relative to most optical setups where only the two-mode beamsplitters that operate between adjacent modes are usually easy to implement.  In the future, other useful beamsplitter constructions may be obtained by considering phase-modulation theory, using phase modulation and single-mode phase shifts to create arbitrary unitary operations.  Investigations into other experimentally convenient tools for linear control will remain an area for further research.

\section{Operational Fidelity}

The fundamental sources of error present in the atomic spin-wave processor are discussed in the joint Letter \cite{cox_spin-wave_2021}.  These errors, present for any implementation using optical qubits in a cavity, arise from atomic saturation and atomic emission into the cavity and into free space.  Here we discuss additional technical sources of error that will likely arise in the cold-atom implementation.

\subsection{Effect of momentum displacement errors}

First we consider the effect of small amplitude errors in the momentum displacement operator $\hat{\Delta}$.  This operator works correctly when the amount of phase shift leads to an integer change in the momentum index.  We consider the effect of small imperfections $\epsilon$ in the amplitude of this  operation, that results in a non-integer momentum $k \rightarrow k + 1 + \epsilon$.  The error $\eta_\Delta$ is calculated to be,
\begin{align}
    \eta_\Delta &= 1-|\!\langle b_{k+\epsilon} | b_k\rangle |^2 \notag\\
    &= 1-\bigg|\frac{1}{M} \sum_{j=1}^{M} e^{2 \pi i \epsilon j / M}\bigg|^2\notag\\
    &\approx \pi^2 \epsilon^2,
\end{align}
in the limit of small $\epsilon$ and large $M$. The loss of quantum fidelity is second order in the error $\epsilon$.  But nonetheless, the $\hat{\Delta}$ operation will require good amplitude control.  More complex pulse sequences that are amplitude independent to higher order---similar to those used in Nonlinear Magneto-Optical Rotation (NMOR), pulsed spectroscopy, dynamic decoupling, and optimal control \cite{khaneja_optimal_2005, rakreungdet_accurate_2009, glaser_training_2015}---may be useful to eliminate this error in experimental settings.

\subsection{Errors from variation in atom number}
The goal of this apparatus will be to achieve approximately constant atom number per site.  However, some variation will likely remain.  We estimate the errors from this variation.  In the case of non-uniform atom number per site, the cavity dressing interaction is re-written using a non-uniform projector $\ket{b'_0}$
\begin{align}
    H^{k*}_0 &= \frac{\hbar\Omega_2^2}{4 \delta_2} \ket{b_0'}\bra{b_0'},\\
    \ket{b_0'}&= \sum_{j=0}^{M-1}  \sqrt{\frac{n'_j}{N}}  \ket{a_j},\\
    \ket{a_j} &= \frac{1}{\sqrt{n_j'}} \sum_{l = 0}^{n'_j-1} \ket{g_l}\!\bra{e_l},
\end{align}
where $n'_j = n(1+\epsilon_j)$ is the erroneous factor describing the non-uniform atom number at site $j$ and the errors $\epsilon_j$ are assumed to sum to zero.  With atom number variation, this cavity coupling Hamiltonian is not equivalent to $H^k_0 \propto \ket{b_0}\!\bra{b_0}$.  The cavity interaction leads to a phase shift in a new mode $\ket{b_0'}$, a mode that is not trivially decomposable into the orthogonal basis.  Assuming the orthogonal basis must be maintained for the desired operations, this leads to an error of
\begin{align}
\eta_N &= 1 - |\!\braket{b_0|b_0'}\! |^2 \notag\\
&\approx \frac{1}{M} \sum_{j=0}^{M-1} \frac{\epsilon_j^2}{4}
\end{align}
 for small errors $\epsilon_j$.  $\eta_N$ will likely be dominated by static inhomogeneities in atom number, and dealing with non-uniform ensembles may require additional work in the future.  These static errors may be correctable using compensation techniques in the pulses or perhaps appropriate re-definition of the basis.

\subsection{Readout}
Another important ingredient for the spin-wave quantum processor is readout.  Readout is required for almost all photonic processes, and is necessary for linear optical quantum computing.  The readout process is not a focus of this manuscript, because spin-wave readout has been studied in-depth by many previous quantum memory experiments \cite{choi_entanglement_2010, simon_single-photon_2007, yang_efficient_2016}.  In particular, atom-cavity systems demonstrate the most efficient readout of any type of quantum memory with intrinsic readout probabilities of well over 90\% possible \cite{yang_efficient_2016}.  Many effective quantum networking protocols are stable to inefficiencies at this level \cite{sangouard_quantum_2011}.

\section{Initialization time}

\begin{figure}[t]
\centering
\includegraphics[width=\columnwidth]{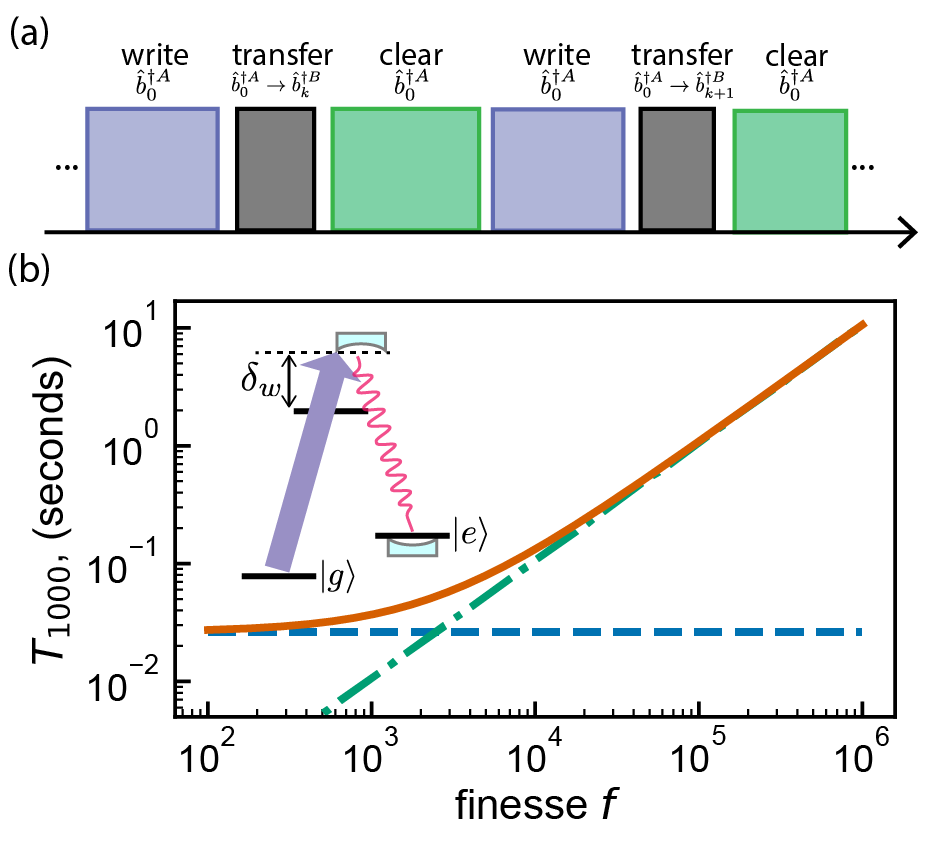}
\caption{(a) Initialization sequence.  Excitations are written into ensemble $A$ and transfered to ensemble $B$ to prevent compounding errors.  (b) Approximate initialization time for 1000 single-photon excitations in the ensemble for a 2~cm length optical cavity. The initialization time is limited by both the cavity lifetime (green dot-dashed, given by $1000/(\eta_w \kappa$)) and atomic excited state lifetime (blue dashed, given by $1000/(\eta_w\Gamma$)) with the total time shown in orange (solid).  The level diagram for memory initialization is shown in the inset.}\label{fig:init} 
\end{figure}

Initialization of single-photon excitations in the spin-wave memory may be achieved by several different methods.   Here, we consider probabilistically creating, in rapid succession, single excitations in a large array of momentum eigenstates.  The level diagram for this write process is shown in the inset of Fig.~\ref{fig:init}(b).  The level scheme is the inverse of the diagram in Fig.~\ref{fig:expt} and requires one additional longitudinal cavity mode, that can easily be selected with the frequency of the initialization laser (also called the write laser).  The initialization laser must counter-propagate relative to the dressing laser to maintain phase matching of both the read and write photons into the cavity mode \cite{simon_interfacing_2007}.

Memory initialization is accomplished using the standard atomic memory heralded write process into the $\ket{b_0}$ mode \cite{duan_long-distance_2001}, followed by a unit displacement $\hat{\Delta}$, repeated until a large fraction of the spin-wave modes are initialized.    It is important to keep the probability of double excitation low, since such errors are not detected by heralding.  In atomic memories, the double excitation error $\eta_2$ is proportional to the write probability $p_1$, $\eta_2 \propto p_1$, meaning that the write probability must be kept small \cite{sangouard_quantum_2011}.  However, when excitations are initialized within an $M$-mode register, the standard double-excitation error $\eta_2$ due to atomic emission into free space is amplified.  Normally, the full error from the write process $\eta_w$ is of scale $\eta_2$  \cite{sangouard_quantum_2011}.  However, in an M-mode spin-wave register, each mode gains an independent error of scale $\eta_2$ for every write process, so that the total error in each mode compounds to a larger value $\eta_w \sim M \eta_2$.

In order to overcome this unfortunate scaling, we propose a modified heralded initialization scheme  that works in the two-ensemble configuration.  The pulse sequence is displayed in Fig.~\ref{fig:init}(a).  By writing excitations into a single ensemble (chosen to be mode $\hat{b}_0^A$ here), the excitation can be initialized and transferred with a low-error beamsplitter operation into ensemble $B$.  Ensemble $A$ can be cleared with a standard optical pumping pulse (labeled ``clear''), before subsequent excitations are written.  The optical pumping prevents errors in the initial write procedure from compounding in later steps.

In Fig.~\ref{fig:init}(b), we plot the approximate initialization time required to initialize 1000 modes.  The speed limits for the memory write process are dictated by the excited state linewidth $\Gamma$ and the cavity linewidth $\kappa$.  Using these rates, and maintaining a single write error $\eta_w \sim \eta_2 \sim p_1$ of less than 0.001, we plot the estimated time $T_{1000}$ required to create 1000 single excitations in Fig.~\ref{fig:init} versus cavity finesse $f$:
\begin{equation}
T_{1000} \sim \frac{1000}{\eta_w}(\frac{1}{ \Gamma } + \frac{1}{\kappa}).
\end{equation}
The cavity linewidth $\kappa$ is related to finesse by $\kappa = 2 \pi  c/(l f)$ where $c$ is the speed of light and $l$ is the round-trip cavity length.  We see that the cavity lifetime becomes the limiting factor at a finesse of around 2000, for a cavity length of 2~cm, assuming ideal detection efficiency.    Although the initialization time for 1000 excitations and  $f=10^4$ is still well below the maximum atomic lifetimes observed in spin-wave memories \cite{dudin_light_2013}, this speed limit may  be a significant concern for future high capacity memories in high finesse cavities. Deterministic initialization methods involving single photon sources or Rydberg excitations may be necessary to consider in the future \cite{li_quantum_2016, ornelas-huerta_-demand_2020}.

\section{Squeezing and continuous variable processing}
Atom-cavity experiments have recently generated record amounts of squeezing, entanglement useful for enhancing quantum sensors \cite{hosten_measurement_2016, cox_deterministic_2016}.  Squeezing of 20~dB, or a factor 100 in variance, is now achievable in systems similar to the one proposed here.  Additionally, recent optical experiments have shown how to use squeezing as a powerful computational resource to create dual-rail cluster states \cite{asavanant_generation_2019}.  
Cluster states are particularly appealing for future quantum processors because they are deterministic and are amenable to powerful continuous variable quantum error correction schemes \cite{fukui_high-threshold_2018, bourassa_blueprint_2021}.  
By creating an $M$-mode processor, we now open the possibility to combine the cluster state concept with large amounts of atomic spin squeezing to build a continuous variable atomic processor.  
Here, we describe how to implement the optical dual-rail cluster state scheme in the atom-cavity system.  The cold atom system leads to numerous advantages relative to the photonic implementation since the qubits are stationary and amenable to feedback and high-fidelity processing with no losses.  

There are several viable paths to create spin squeezing in the spin-wave memory including one- and two-axis twisting and quantum non-demolition (QND) measurements.  For example, the squeezing operation may be described by an operator $\hat{S}(\alpha) = \exp (\frac{1}{2} (\alpha \hat{b}_0^2 - \alpha \hat{b}_0^{\dagger 2}))$ that squeezes only the $k=0$ mode.  In this case, the mode operators are transformed as 
\begin{align}
\hat{b}_0  &\rightarrow \hat{S}^\dagger(\alpha) \hat{b}_0 \hat{S}(\alpha) = \hat{b}_0 \cosh (\alpha) -  \hat{b}_0^\dagger \sinh (\alpha),\\
\hat{b}_k &\rightarrow \hat{S}^\dagger(\alpha) \hat{b}_k \hat{S}(\alpha) = \hat{b}_k \,\,\, (k\neq 0).
\end{align}
Critically, 
the operator $\hat{S}$ does not affect modes $\hat{b}_k$ with $k\neq0$, that are orthogonal to $\hat{b}_0$.  This operation allows us to create independently squeezed spin-wave modes.   In the next section, we present a description of how to create an $M$-mode squeezed state where each $k$-mode is spin squeezed.  This state, when passed through $2M-1$ beamsplitters, transforms into a dual-rail cluster state that may be used for universal quantum computation.  This method is a direct adaptation of seminal results in the optical regime, creating continuous variable cluster states of light \cite{yokoyama_ultra-large-scale_2013}.

\section{Cluster state generation}

Spin-wave continuous-variable quantum computing can be achieved using the same basis as linear optical spin-wave computing.  We consider a similar protocol to experiments in the optical regime that have generated dual-rail cluster states with over 10000 nodes \cite{yokoyama_ultra-large-scale_2013}. 

A continuous-variable cluster state is a large entangled state defined by nullifiers, analagous to the stabilizers of a discrete cluster state \cite{menicucci_universal_2006}.  The nullifiers are joint operators that describe noise projection of nearest-neighbor spin-waves.  The nullifiers in our spin wave case are,
\begin{align}
    \epsilon^x_k &= \hat{\mathcal{X}}^A_k + \hat{\mathcal{X}}^B_k+\hat{\mathcal{X}}^A_{k+1}-\hat{\mathcal{X}}^B_{k+1},\\
    \epsilon^p_k &= \hat{\mathcal{P}}^A_k + \hat{\mathcal{P}}^B_k-\hat{\mathcal{P}}^A_{k+1}+\hat{\mathcal{P}}^B_{k+1},
\end{align}
 for any mode $k$ (modulo $M$) where $\hat{\mathcal{X}}^A_k$ and $\hat{\mathcal{P}}^A_k$ are the quadrature operators for the $\hat{b}^A_k$ spin wave.  Optical experiments have achieved entanglement, as detected by a reduction in $|\!\braket{\epsilon^x}\!|^2$ and $|\!\braket{\epsilon^p}\!|^2$ below a value of 1/2.  Current demonstrations have reached values of around $-6$ dB \cite{yokoyama_ultra-large-scale_2013}.  State-of-the-art spin squeezing may be able to reduce these quadrature values to $-20$ dB or smaller, potentially reaching the fault tolerant threshold for GKP-type quantum error correction \cite{tzitrin_progress_2020, bourassa_blueprint_2021}.

\begin{figure}[t]
\centering
\includegraphics[width=\columnwidth]{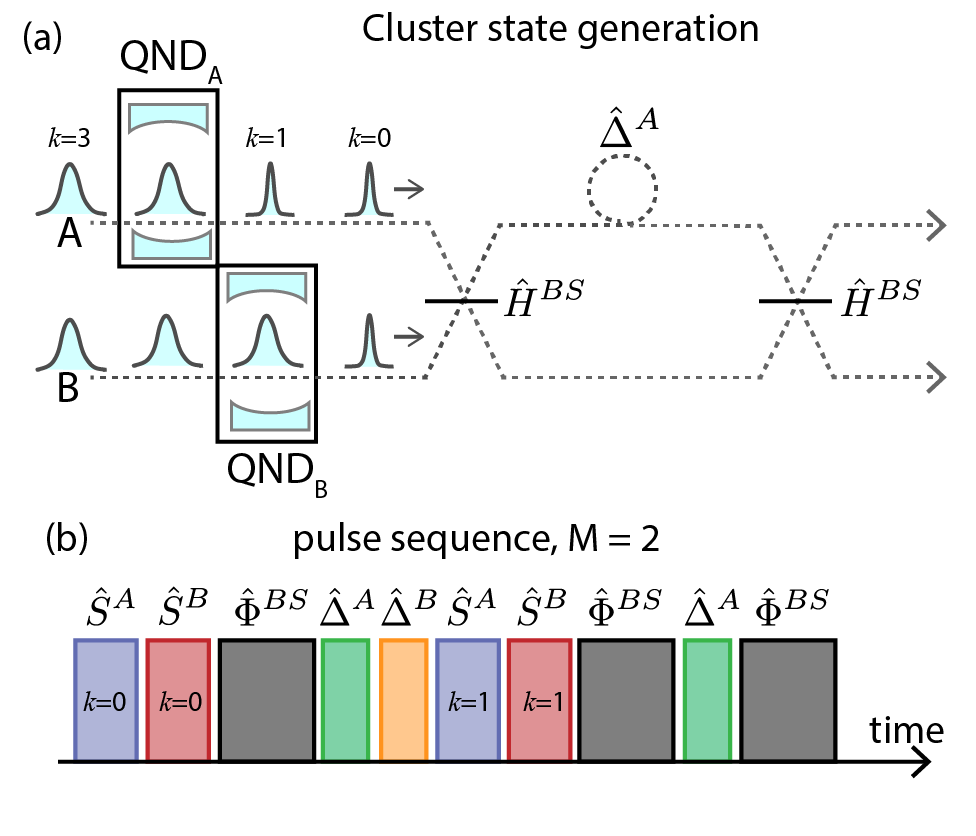}
\caption{Cluster state generation scheme. (a) Optical diagram for generating a spin-wave cluster state.  QND measurements are sequentially applied to squeeze each momentum state.  Using two beamsplitters and a phase shift $\hat{\Delta}^A$, a dual rail cluster state is created.  Additional unshown momentum shift operators $\hat{\Delta}^A$ and $\hat{\Delta}^B$ are necessary to transfer spin waves in and out of the $k=0$ interaction mode.  (b) Example pulse sequence for generating a dual-rail cluster state with $M = 2$.  Pulse colors are set to match Fig.~\ref{fig:expt} and the squeezing pulses are labeled with the affected initial value of $k$.} \label{fig:fig4}
\end{figure}

An experimental diagram is shown in Fig.~\ref{fig:fig4}.  Two ensembles (labeled A and B) are loaded into the bowtie cavity, one on each side.  The ensembles each provide a basis of momentum states $\ket{b^{A}_k}$ and $\ket{b^{B}_k}$ for $0 \leq k <M$.  The goal is to use collective cavity quantum non-demolition (QND) measurement or other technique \cite{hosten_measurement_2016, cox_deterministic_2016} to generate spin-squeezing in each momentum mode, and then use collective cavity interactions to emulate the beamsplitters required to transform the squeezed modes into a cluster state.

First, cavity QND measurements or other squeezing operations $\hat{S}$ are performed on each individual spin wave to create a stream of squeezed states.  Spin waves are transferred into and out of cavity coupling with the $\hat{\Delta}^A$ and $\hat{\Delta}^B$ operators.  Using the well-known construction for creating dual-rail cluster states \cite{yokoyama_ultra-large-scale_2013}, two-mode beamsplitters $\hat{H}^{BS}$ and phase shifts $\hat{\Delta}^A$ and $\hat{\Delta}^B$ are sufficient for creation.  Unlike optical cluster states, the atomic dual rail cluster state is stationary with long coherence time, and hence amenable to real-time computation.

\begin{figure}[t]
\centering
\includegraphics[width=\columnwidth]{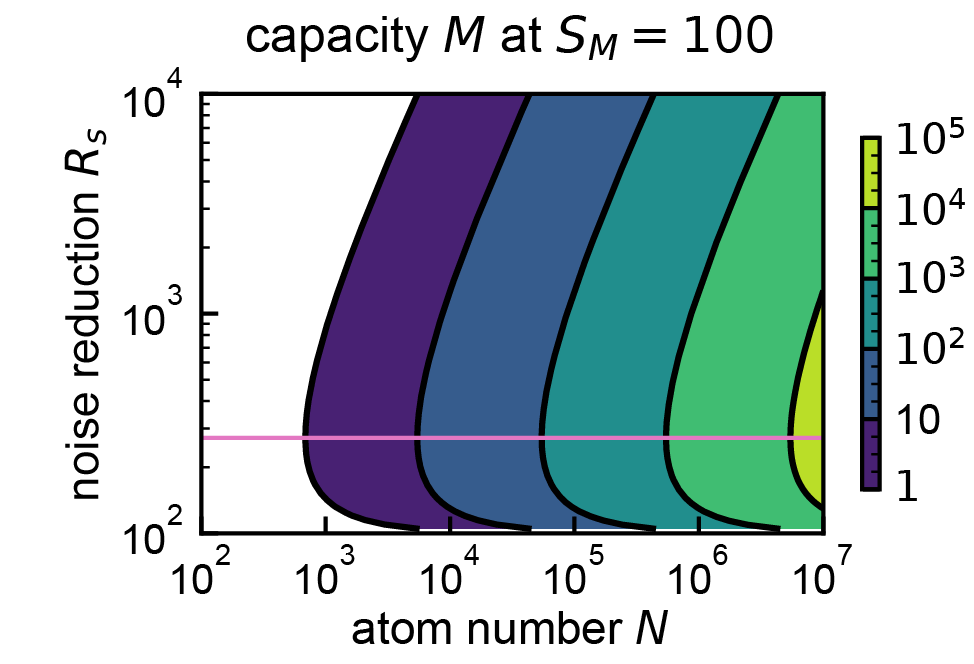}
\caption{Capacity $M$ of a spin-wave cluster that can achieve 20~dB of spin squeezing simultaneously in each mode ($S_M = 100$) as a function of the atom number $N$ and squeezing $R_s$, the quantum noise reduction in each mode.  $M$ is optimized at $R_s = e S_M$ (pink line)}\label{fig:cluster}
\end{figure}

In Fig.~\ref{fig:cluster}(a) and (b), we estimate the maximum capacity $M$ of the spin-wave processor for storing a large cluster state.  The amount of squeezing and the capacity will be limited by several factors, including the beamsplitter errors \cite{cox_spin-wave_2021} and the ability for the atom-cavity system to generate squeezing in the first place.  But one fundamental limitation is the capacity of the atomic system to store spin-squeezed states without a subsequent reduction in coherence due to nonlinearity (i.e., curvature of the Bloch sphere leading to nonlinear projections of the Bloch vector onto a 2-D plane).

State-of-the-art atom-cavity experiments can generate spin squeezing with spectroscopic enhancement near a factor of one hundred \cite{hosten_measurement_2016,cox_deterministic_2016}.  Spectroscopic enhancement, or amount of squeezing, is defined as the entanglement-generated improvement in the sensor's ability to resolve a quantum phase.  For a single spin wave, we write the spectroscopic enhancement  \cite{chen_cavity-aided_2014}
\begin{equation}
    S_s = R_s \mathcal{C}^2,
\end{equation}
where $R_s =2\textrm{Var}(\hat{J}_z)/N$ is the reduction in variance of the expectation value of the collective spin operator $\hat{J}$ along a particular axis (chosen as $z$ here).   $\mathcal{C}$ is the spin coherence of the ensemble defined as $\mathcal{C} = J/(N/2)$, where $J \equiv \langle \hat{J} \rangle$ is the expectation value of the total projection of $\hat{J}$.  $J$ can have values between $N/2$ (full spin coherence) and $0$ (no spin coherence).  $S_s>1$ is both a witness for atomic entanglement as well as a measure of the entanglement-generated improvement in the quantum sensor.
  
We now consider simultaneous equal squeezing in all $M$ modes of an ensemble.  In this case, we reduce the quantum noise in all spin-wave modes by an equal amount $R_s$.  $S_s$ is defined to be the squeezing that would be observed in a spin-wave, if no other modes were squeezed.  However, when all modes are squeezed at 
the same time, we observe a lower amount of squeezing in each mode, that we denote $S_M$.   
To calculate $S_M$, we  must include the additional reduction in total $\mathcal{C}$ due to Bloch sphere curvature, that leads to a compounding reduction in squeezing in every mode.  Then, the observed squeezing of a single mode in the presence of squeezing in all other modes is
\begin{align}\label{eq:sm1}
    S_M &\approx R_s \prod_{j=0}^{M-1} C_j^2, 
\end{align}
where $C_j$ is the spin coherence in each mode that additionally limits the total squeezing.  The coherence of each mode is limited by quantum back action and the curvature of the Bloch sphere \cite{andre_atom_2002}.  Each spin-wave has an rms (root mean squared) back action around the Bloch sphere of angle $\theta_{rms} = \theta_{SQL} \sqrt{R_s}$ where $\theta_{SQL} = 1/\sqrt{N}$ is the standard quantum limit in radians (see, for example, Ref.~\cite{cox_deterministic_2016}).  This leads to a fundamental loss in coherence in each spin wave, due to Bloch sphere curvature:
\begin{align}\label{eq:contrast}
    C_j &\approx e^{-\theta_{rms}^2/2} \approx e^{-\theta_{SQL}^2 R_s/2}.
\end{align}
By combining Eq.~(\ref{eq:contrast}) and Eq.~(\ref{eq:sm1}), the maximum $M$-mode squeezing is found to be
\begin{align}
    S_M &\approx R_s e^{-  M \theta^2_{SQL} R_s }.
\end{align}
 Quantum inefficiency and other decoherence sources that would cause further loss of coherence are not considered.

In Fig.~\ref{fig:cluster} we plot the value of $M$ that is achieved for $S_M= 100$, that is, 20~dB of squeezing versus $N$ and $R_s$.  The maximum capacity $M_{opt}$ is found at $R_s = e S_M$.  The maximized value is 
\begin{equation}
    M_{opt} = \frac{N}{2 e S_M}.
\end{equation}
A processor reaching the levels of performance in Fig.~\ref{fig:cluster} would be state-of-the-art and likely useful for deterministic quantum networking and entanglement distribution.  Further, 20~dB of squeezing in each mode would reach or nearly reach the fault-tolerant threshold \cite{bourassa_blueprint_2021} for continuous variable quantum computing.  

\section{Near-term Applications}
Overall, deterministic continuous-variable quantum computing has significantly better prospects for scalability than linear optics with single-quanta excitations, because the linear optics scheme requires a significant resource overhead to achieve deterministic processing \cite{bourassa_blueprint_2021}.  However, in the near term, small-scale linear optical processing inside of a multiplexed quantum memory may be a significant boon toward realizing a quantum repeater with medium to high speed of entanglement generation over long distances (over 100~km) \cite{cox_spin-wave_2019,muralidharan_optimal_2016}.  Meanwhile, in the continuous-variable scheme, the creation of spin-wave entangled states and the study of their uses for quantum sensing and networking applications should be one of the first experimental goals.

The multimode quantum processor presented here is also ideal for certain classes of quantum sensing problems that involve data distributed between multiple modes.  The spin wave processor utilizes $2M$ independent modes within the two $N$-atom ensembles, yielding enhanced capability for certain classes of measurements.  Recent work has shown that distributed quantum sensors enable new sets of applications involving measurements of extended systems \cite{zhuang_distributed_2018,qian_heisenberg-scaling_2019, guo_distributed_2020, qian_optimal_2021}, and the spin wave processor may extend these protocols to sensing and receiving data distributed into multiple spatial or temporal modes.

\section{Technical Challenges}\label{sec:challenge}
Additional technical challenges will likely arise when building an experiment to accomplish this proposal.  The application of the Raman dressing laser results in a shift in the cavity resonance frequency by amount $\Delta \omega_c \approx N_g g^2 / \delta_1$ and a shift in the two-photon resonance frequency $\omega_{eg}$ by amount $\Delta \omega_{eg} \approx \Omega_d^2/\delta_1$.  These shifts will need to be taken into account to achieve accurate dynamics in the two-photon system. To achieve effective operations, the laser detunings may need to be actively stabilized to account for these cavity and state shifts, as has been done in recent entanglement-generation experiments \cite{hosten_measurement_2016, cox_deterministic_2016}.  

Rapidly addressing a large capacity of quantum bits or modes is a ubiquitous challenge in nearly every experimental quantum platform.  Here, we achieve that end simply, using a single optical cavity mode and an off-resonant light source.  One advantage of the proposal is that the $\hat{\Delta}$ operator can be implemented with light that is at a wavelength far from atomic resonance, with no stringent wavelength or power requirements.  We envision a fast electro-optic system operating in the near-infrared, which can operate with switching ranges of well over 1~GHz.

\section{Conclusion}

Overall, we are optimistic that holographic spin-wave excitations in a cavity-coupled ensemble may become a useful platform for quantum information processing.  The system combines several attractive characteristics including efficient readout into a single optical cavity mode, large capacity, and universal processing capabilities able to achieve high fidelity.  In the long term, many quantum information processing devices will likely require networked operation, and atom-cavity systems will be a preeminent platform to achieve this.

\section{Acknowledgements}

P.B.~and A.V.G.~acknowledge funding by ARO MURI, DARPA SAVaNT ADVENT, AFOSR MURI, AFOSR, NSF PFCQC program, DoE ASCR Accelerated Research in Quantum Computing program (award No.~DE-SC0020312), the DoE ASCR Quantum Testbed Pathfinder program (award No.~DE-SC0019040), and U.S.~Department of Energy Award No.~DE-SC0019449.  

\label{app:ExpDetails}

\bibliography{bib1}

\end{document}